\documentclass[aps,prl,twocolumn]{revtex4-1}
\usepackage{graphicx}
\usepackage{color}

\makeatletter
\@ifundefined{textcolor}{}
{
 \definecolor{BLACK}{gray}{0}
 \definecolor{WHITE}{gray}{1}
 \definecolor{RED}{rgb}{1,0,0}
 \definecolor{GREEN}{rgb}{0,1,0}
 \definecolor{BLUE}{rgb}{0,0,1}
 \definecolor{CYAN}{cmyk}{1,0,0,0}
 \definecolor{MAGENTA}{cmyk}{0,1,0,0}
 \definecolor{YELLOW}{cmyk}{0,0,1,0}
}

\begin{document}
\title{Enhancement of the superconducting gap by nesting in CaKFe$_4$As$_4$ - a new high temperature superconductor}
\author{Daixiang Mou$^{1, 2}$, Tai Kong$^{1, 2}$, William R. Meier$^{1, 2}$, Felix Lochner$^3$, Lin-Lin Wang$^{2}$, Qisheng Lin$^{1}$, Yun Wu$^{1, 2}$, S. L. Bud'ko$^{1, 2}$, Ilya Eremin$^3$, D. D. Johnson$^{1, 2, 4}$, P. C. Canfield$^{1, 2}$, Adam Kaminski$^{1, 2}$}
\affiliation{
\\$^{1}$Division of Materials Science and Engineering, Ames Laboratory, Ames, Iowa 50011, USA
\\$^{2}$Department of Physics and Astronomy, Iowa State University, Ames, Iowa 50011, USA
\\$^3$Institut fur Theoretische Physik III, Ruhr-Universitat Bochum, 44801 Bochum, Germany
\\$^4$Department of Materials Science and Engineering, Iowa State University, Ames, Iowa 50011, USA
}

\begin{abstract}
We use high resolution angle resolved photoemission spectroscopy and density functional theory with experimentally obtained crystal structure parameters to study the electronic properties of CaKFe$_4$As$_4$. In contrast to related CaFe$_2$As$_2$ compounds,  CaKFe$_4$As$_4$ has high T$_c$ of 35K at stochiometric composition. This presents unique opportunity to study properties of high temperature superconductivity of iron arsenic superconductors in absence of doping or substitution. The Fermi surface consists of three hole pockets at $\Gamma$ and two electron pockets at  the $M$ point. We find that the values of the superconducting  gap are nearly isotropic, but significantly different for each of the FS sheets. Most importantly we find that the overall momentum dependence
{of the gap magnitudes plotted across the entire Brillouin zone displays a strong deviation from the simple $\cos(k_x)\cos(k_y)$ functional form of the gap function, proposed in the scenario of the Cooper-pairing driven by a short range antiferromagnetic exchange interaction. Instead, the maximum value of the gap is observed for FS sheets that are closest to the ideal nesting condition in  contrast to the previous observations in some other ferropnictides. These results provide strong support for the multiband character of superconductivity in CaKFe$_4$As$_4$, in which Cooper pairing forms on the electron and the hole bands interacting via dominant interband repulsive interaction, enhanced by FS nesting}.
\end{abstract}
\pacs{74.25.Jb, 74.72.Hs, 79.60.Bm}
\maketitle

Understanding the superconducting mechanism in iron-based, high temperature superconductors is an important  topic in condensed matter physics. One of the key questions is  whether the system should be described within a
{weak coupling BCS-type approach with the key role played by the interband repulsion between electron and hole bands, separated by the large momentum transfer} or by a strong
{coupling approach with dominant short-range antiferromagetic fluctuations described by the local exchange interaction}\cite{Chubukov2008,Mazin2009,Graser2009,Si2016,Hu2012,Inosov2011,Richard2015}. The former scenario seemed consistent with experimental results from number of iron pnictide superconductors\cite{Richard2011}, but was challenged later by the discovery of iron chalcogenide superconductors\cite{Mou2011a,Liu2012}. The theoretical progress made in this field was normally inspired by the discovery of new materials in this family \cite{Kamihara2008} from iron chalcogenide\cite{Guo2010} to single layer FeSe film\cite{Qing-Yan2012}. Materials with different crystal or electronic structures are always extremely useful to provide new insights and constructing global model of high temperature superconductivity iron based materials.

Recently, a new Fe-based class of superconductors, \emph{AeA}Fe$_4$As$_4$ (\emph{Ae} = Ca, Sr, Eu and A = K, Rb, Cs), were reported (generically referred to as $AeA1144$)\cite{Iyo2016,2016Meier}. Although the chemical composition of $AeA1144$ is the same as the intensively studied (Ba, K)Fe$_2$As$_2$ superconductors, it has a different crystal structural type with \emph{Ae} and A layers alternatively stacked between Fe$_2$As$_2$ layers. The crystallographically inequivalent position of the \emph{Ae} and A  atoms changes the space group from \emph{I4/mmm} to \emph{P4/mmm}. Furthermore,  different valence attraction from \emph{Ae}$^{2+}$ and A$^{1+}$  layers to Fe$_2$As$_2$$^{1.5-}$  and ionic radius leads to different length of As-Fe bonds, which was proposed as an important parameters controlling T$_c$ of Fe-based superconductors\cite{Mizuguchi2010}. Together with its high transition temperature (T$_c$ = 31 - 36 K), the $AeA1144$ family provides a new platform to test existing theories and inspire new ones.  Measurements of its electronic structure and momentum dependence of the superconducting gap are of critical importance.

In this letter, we investigate the band structure and momentum dependence of the superconducting gap of CaKFe$_4$As$_4$ superconductor with high resolution Angle Resolved Photoemission Spectroscopy (ARPES) and DFT utilizing experimentally obtained crystal structure parameters. Unlike most other Fe-based superconductor, CaKFe$_4$As$_4$ has a high T$_c$ of 35 K at stoichiometric composition which allows for the study of iron based high temperature superconductivity in absence of substitution introduced disorder. We find that the linewidths to be relatively broad and comparable to substituted Ba(Fe, Co)$_2$As$_2$ signifying that relatively large scattering is intrinsic and not a result of substitution disorder. This settles a long dispute about the role of substitution disorder \cite{Sawatzky}. The Fermi surface consists of three hole pockets at the $\Gamma$ point and two electron pockets at the $M$ point. The size of the hole pockets is significantly different, which allows us to measure the superconducting gap for different values of the total momentum {\bf k}. We find that the values of the superconducting  gap  are nearly isotropic, but significantly different for each of the Fermi surface(FS) sheets. Indeed, the largest superonducting gap is found for pairs of hole and electron pockets that have similar diameter, while other pockets have smaller value of the superconducting gap.
{This observation is in stark contrast to the situation in some other ferropnictides. For example, in LiFeAs, which is another stoichiometric pnictide superconductor with multiple Fermi surface sheets, the largest superconducting gap was found to be on the smallest hole pocket, located near the zone center.\cite{Borisenko2012,Umezawa2012} Therefore, our results on CaKFe$_4$As$_4$ provide new important ingredient the superconducting mechanism of Fe-based superconductor must contain}.

\begin{figure}[htbp]
\centering
\includegraphics[width=0.8\columnwidth]{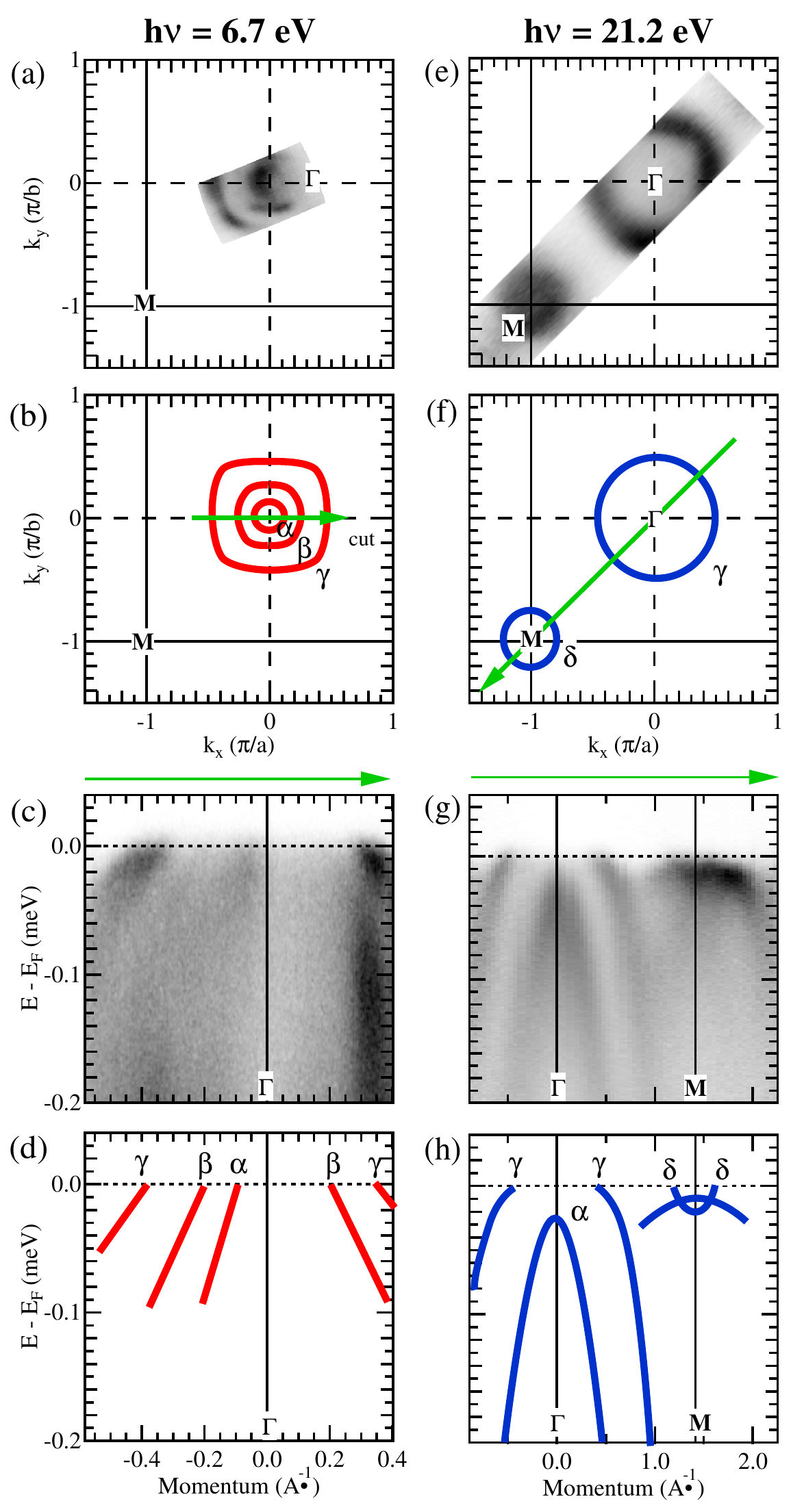}
\caption{Measured electronic structure of CaKFe$_4$As$_4$. (a) Fermi surface intensity acquired using photon energy of 6.7 eV and T=40 K. (b) Sketch of the FS based on data in (a). (c) Measured ARPES intensity along a cut through $\Gamma$ point. Cut position is indicated in panel (b). (d) Sketch of the band structure based on data in (c). (e)-(h) Same as (a)-(d), but measured using photon energy of 21.2 eV.}
\end{figure}

\begin{figure}[htbp]
\centering
\includegraphics[width=0.9\columnwidth]{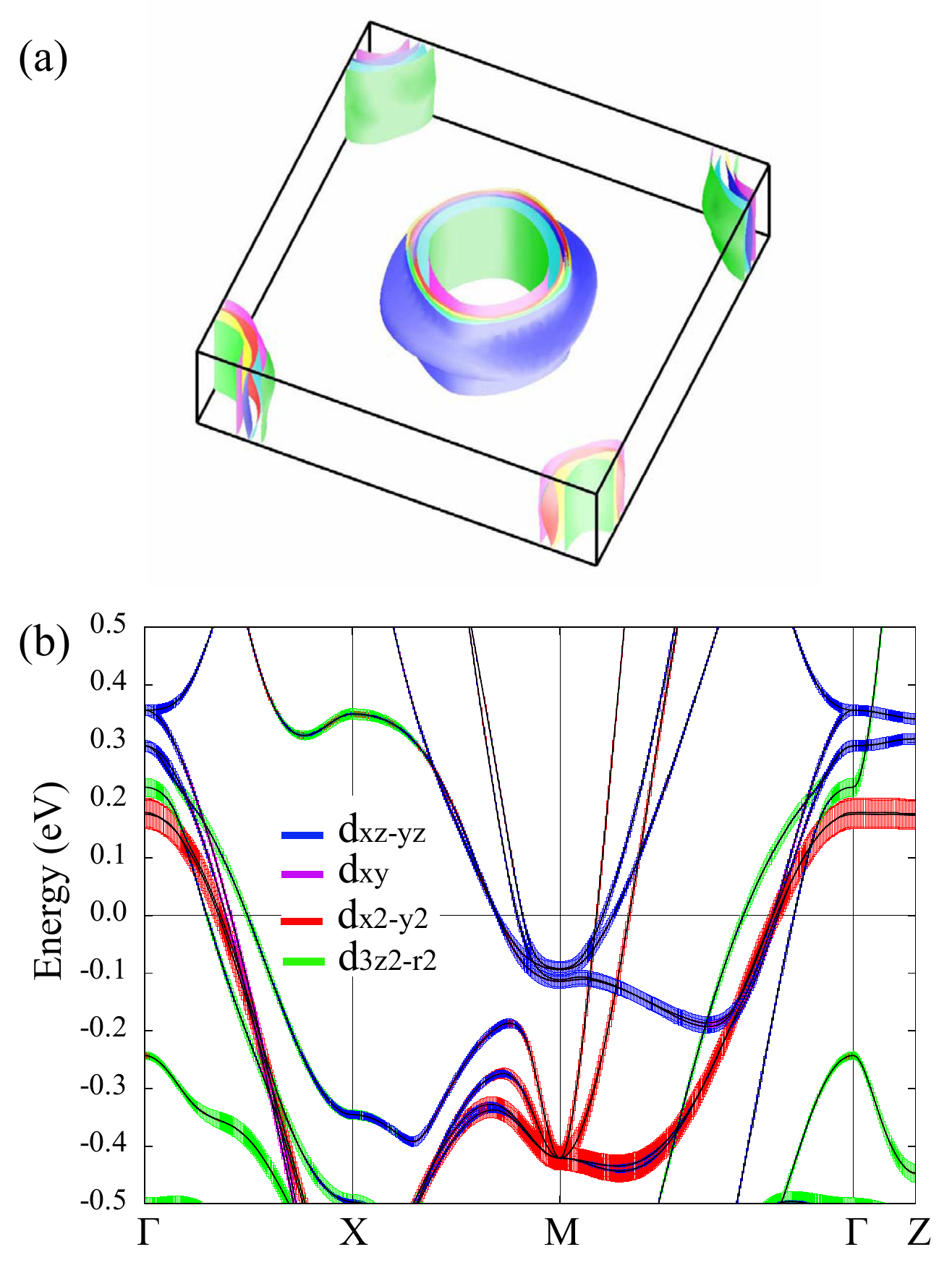}
\caption{(a) calculated 3D Fermi surface of CaKFe$_4$As$_4$ (b-e) band dispersion along key symmetry directions with respective orbital contributions marked by color coded outlines.}
\end{figure}

\begin{figure*}[htbp]
\centering
\includegraphics[width=0.9\textwidth]{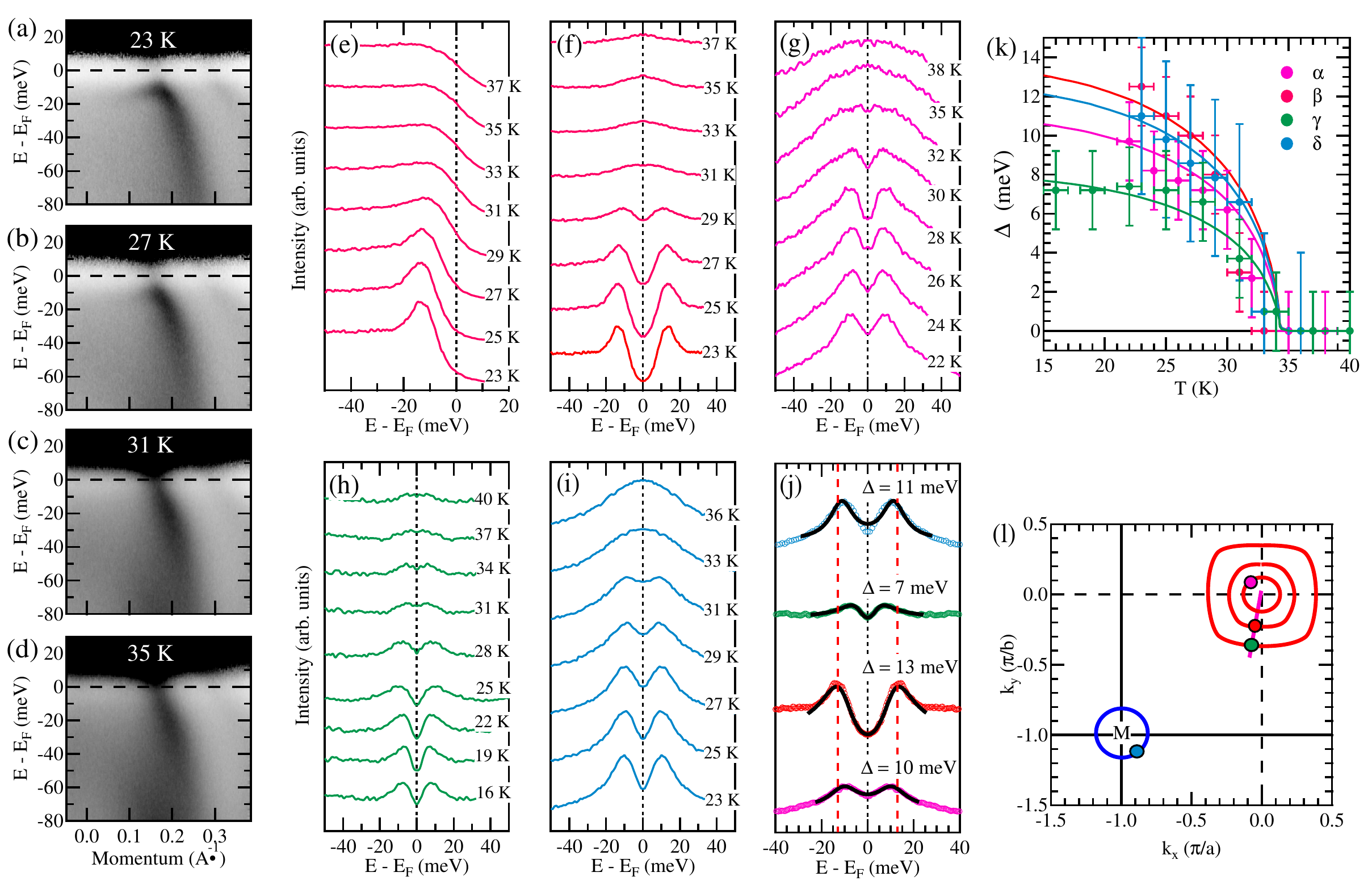}
\caption{Temperature and Fermi sheet dependence of the superconducting gap. (a)-(d) Measured electronic structure at four selected temperatures. The data is divided by the Fermi-Dirac function to illustrate the opening of the superconducting gap. Cut position is indicated in (l). (e) EDCs at k$_F$ of $\beta$ FS pocket. (f) The data from panel (e) after symmetrization. (g)-(i) Symmetrized EDCs at k$_F$ of $\alpha$, $\gamma$ and $\delta$ FS pockets respectively. k$_F$ positions are marked in (l). (j) Symmetrized EDCs at k$_F$ for lowest measured temperature from (f)-(i). Red, vertical dashed lines mark the energy of largest gap ($\beta$ FS sheet) for easier comparison. The black lines are the fits using phenomenological model \cite{Matsui2003}. (k) Temperature dependence of the superconducting gaps for all four pockets. Solid lines are BCS predictions for $\Delta_0$ of 10.5 meV, 13 meV, 8 meV and 12 meV. (l) Sketch of the FS with indication of the cut position and k$_F$ positions.}
\end{figure*}

CaKFe$_4$As$_4$ single crystals were grown using flux method and extensively characterized by thermodynamic and transport measurements\cite{2016Meier, 2016Kong}.
The experimental structure parameters were carried out using single crystal X-ray diffraction. Technical details are provided in the Supplemental Material.
 Single phase samples were cleaved \emph{in situ} at a base pressure of lower than 8 $\times$ 10$^{-11}$ Torr. ARPES measurements were carried out using two laboratory-based systems. The electronic structure around the center of the Brillouin zone was measured using laser ARPES system consisting of a Scienta R8000 electron analyzer and tunable VUV laser light source \cite{Jiang2014}. Photon energy was set at 6.7 eV and the energy resolution of the analyzer was set at 4 meV. The electronic structure around the corner of the Brillouin zone was measured with He ARPES system equipped with a Scienta R2002 electron analyzer and microwave plasma Helium lamp ($h\nu$=21.2 eV). The energy resolution of the system was 8 meV. Assuming that the CaKFe$_4$As$_4$  has the same inner potential of (Ba,K)Fe$_2$As$_2$ ($\sim$12 eV), 6.7 eV and 21.2 eV light sources measure the electronic structure around $k_z=\pi$ and $k_z=0$ respectively \cite{Liu2009,Shimojima2011}. Samples were cooled using a closed cycle He-refrigerator and the sample temperature was measured using a silicon-diode sensor mounted on the sample holder. The energy corresponding to the chemical potential was determined from the Fermi edge of a polycrystalline Au reference in electrical contact with the sample. The consistency of the data was confirmed by measuring several samples. 

The Fermi surface and band dispersion were measured by ARPES using two different photon sources with results shown in Fig. 1. In data collected using a photon energy of 6.7 eV, three hole pockets are observed at the center of the zone ($\alpha$, $\beta$ and $\gamma$ shown in panels (a)-(d)). Whereas the $\alpha$ pocket is fairly round, the shapes of the $\beta$ and $\gamma$  pockets are slightly square-ish. The approximate diameters of these three pockets are $\sim$0.2 $\pi/a$, $\sim$0.4 $\pi/a$ and $\sim$0.8 $\pi/a$ respectively.
The FS and band structure data measured using the photon energy of 21.2 eV are shown in Fig. 1(e)-1(h). At this $k_z$ value, only one large hole pocket around $\Gamma$ is observed and its diameter is $\sim$0.45 $\pi/a$. The $\beta$ pocket is not visible, most likely due to unfavorable matrix elements. The band forming $\alpha$ hole pocket, is located 30 meV below E$_F$ for this value of k$_z$ (Fig. 1(g)). At the zone corner, there is a electron FS pocket ($\delta$) from a very shallow band (Fig. 2(g)-2(h) below).

\begin{figure*}[htbp]
\centering
\includegraphics[width=0.8\textwidth]{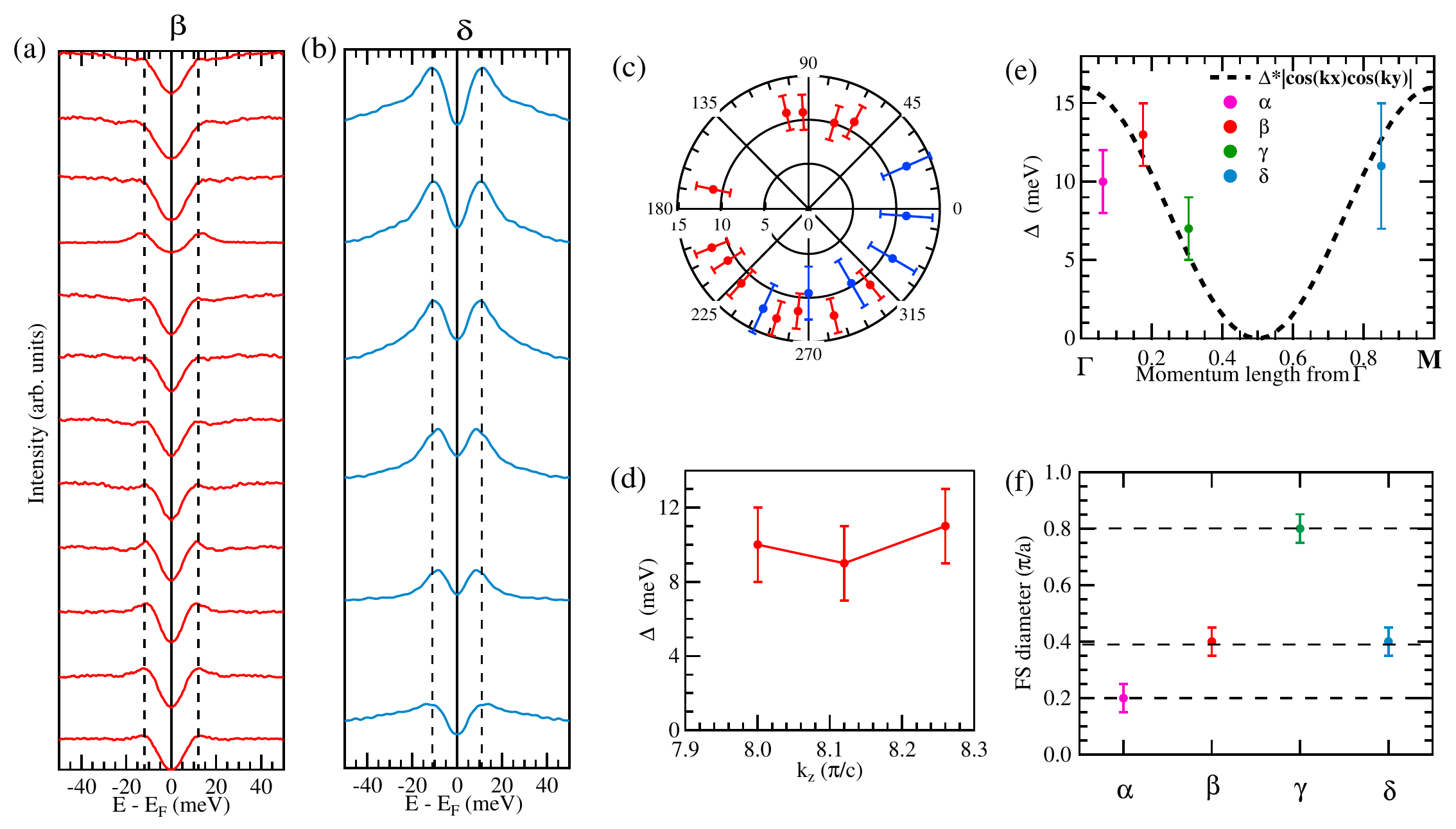}
\caption{Momentum dependent superconducting gap. (a)-(b) Symmetrized EDCs at different K$_F$ on $\beta$ and $\delta$ FS pockets. (c) Extracted superconducting gap from EDCs in (a) and (b). The $\Gamma$-X direction is set to 0 degree.  (d) k$_z$ dependence of superconducting gap on $\beta$ FS pocket. Data were taken with photon energy of 5.7 eV, 6.2 eV and 6.7 eV. Data in (a)-(d) were taken at 22 K. (e) Plot of the extracted gap on four FS in Fig. 3(j) versus momentum distance from $\Gamma$. Black dashed line is a fitting with function of $\Delta|cosk_xcosk_y|$, $\Delta$=16 meV. (f) Measured diameters of four Fermi pockets.}
\end{figure*}

\begin{table}
	\begin{center}
		\begin{tabular}{||c |c |c | c| c |c |c |c ||} 
			\hline
			Atom& Wyck.& site symm. &S. O. F. & x/a & y/b & z/c & Ueq [$\AA^2$]\\ [0.5ex] 
			\hline\hline
			Ca & 1a & 4/mmm& 1 &  0 & 0 & 0 & 0.018(2)\\ [0.5ex] 
			\hline
			K & 1d & 4/mmm & 1 &  1/2 & 1/2 & 1/2 & 0.021(2)\\ [0.5ex] 
			\hline
			Fe & 4i & 2mm & 1 &  0 & 1/2 & 0.7682(2) & 0.012(1)\\ 
			\hline
			As1 & 2g & 4mm & 1 &  0 & 0 & 0.3415(2) & 0.012(1)\\ [0.5ex] 
			\hline
			As2 & 2h & 4mm & 1 &  1/2 & 1/2 & ~0.1231(2)~ & ~0.012(1)~\\ [0.5ex] 
			\hline
		\end{tabular}
	\end{center}
	\label{CrystalStructure}
	\caption{Experimental crystal structure parameters of CaKFe$_4$As$_4$. Space group: P4/mmm (\#123), a=3.8659(12) $\AA$, c=12.884(5) $\AA$. Details are provided in Supplemental Material}
\end{table}


{The Fermi surface and orbitally resolved band dispersion were calculated using experimental lattice constants and atomic positions (obtained from single crystal X-ray diffraction measurements and shown in Table I.) with Local Density Approximation (LDA) in Density Functional Theory (DFT). The resulting FS and band structure are shown in Fig. 2. The electron pockets and the outer hole pockets are slightly deformed, quasi 2D cylinders centered at $\Gamma$ point of the BZ. As in many other iron based superconductors, there are several $3d$ orbitals contributing to the states near the Fermi level. Fig. 2b shows band dispersion with color coded orbital contributions. Most importantly, in CaKFe$_4$As$_4$ in addition to the $yz/xz$ and $x^2-y^2$-orbital contributions to the Fermi surface pockets, there is also a strong admixture of 3$z^2-r^2$-states to the $\alpha$ and $\gamma$ bands, which is somewhat different to the other ferropnictides. In particular, the contribution of 3$z^2-r^2$ orbital to the states near E$_F$ depends sensitively on the Fe ionic positions as the latter are located at the off high-symmetry points in CaKFe$_4$As$_4$. In fact in this system there are two Fe-As distances in the As-Fe-As layer, which is in contrast to the other ferropnictides such as  LiFeAs and CaFe$_2$As$_2$. Furthermore, the $\beta$-band consists of a few bands that are nearly degenerate. }

Fig. 3(a)-3(d) show the measured electronic structure at several typical temperatures along a cut near $\Gamma$ point. We divided the data by the Fermi-Dirac function to better reveal the suppression of spectral intensity  at E$_F$ below T$_c$. Due to matrix elements, only $\beta$ and $\gamma$ bands are seen in this cut. Both bands shows a clear back bending structure at low temperature (Fig. 3(a)) which is regarded as a fingerprint of gap opening. We plot the Energy Distribution Curves (EDCs) at the k$_F$ for several different temperatures in Fig. 3(e). A sharp quasiparticle peak gradually forms as the temperature decreased below T$_c$.  In order to qualitatively analyze the size of superconducting gap, we symmetrized EDCs at k$_F$ as shown in Fig. 3(f). This is a preferred approach to division, as it limits the noise present above E$_F$ that interferes with fitting procedure \cite{Norman1998,Kondo2009}. The data for the other pockets is shown in Fig. 3(g)-3(j).
 A BCS based phenomenological model \cite{Matsui2003} is used to fit the EDCs and extract the gap size. The quality of fits is demonstrated in Fig. 3(j) along with values of extracted superconducting gaps.We repeated this procedure for each pocket at several temperatures with results shown in Fig. 3(k).  All gaps follow BCS-like temperature dependence and close at T$_c$ with $\Delta_{\alpha0}$ = 10.5 meV, $\Delta_{\beta0}$ = 13 meV, $\Delta_{\gamma0}$ = 8 meV and $\Delta_{\delta0}$=12 meV, which give rise to the ratio of 2$\Delta_0$/k$_B$T$_c$ of 7.4, 9.1, 5.6 and 8.4, indicating the superconductivity of CaKFe$_4$As$_4$ is in strong coupling regime.

In order to obtain the information of the  symmetry of the order parameter in CaFeK$_4$As$_4$, we measured the momentum dependence of the  superconducting gap on $\beta$ and $\delta$ FS sheets, as summarized in Fig. 4.   For qualitative analysis, we  extract EDCs at different k$_F$ on the $\beta$ and $\delta$ FS sheets and symmetrise them in Fig. 4 (a) and 4(b). All the symmetrized EDCs show a clear dip structure at E$_F$ and the energy positions of the quasiparticle peaks  do not show much variation with the FS angle. In Figs. 3(c)-(d), we plot the extracted values of the superconducting gap as a function of FS angle. The gap sizes on these two FS pockets have no clear nodes and are roughly isotropic, which directly exclude the possibility of d-wave paring symmetry in CaFeK$_4$As$_4$ superconductor. In order to check the  k$_z$ dependence of the superconducting gap, we measured the gap size on $\beta$ FS with three different photo energies which cover more than 0.2$\pi$/c in momentum space as shown in Fig. 4(e). The gap size does not show much variation within this k$_z$ momentum range either, indicating the quasi-2D nature of CaFeK$_4$As$_4$. 

With the measured amplitude of superconducting gap on all four FS of CaFeK$_4$As$_4$, we can now check the validity of the previous proposed gap function. 
{In the strong coupling approaches\cite{Hu2012b,Yin2014}, the pairing of electrons is introduced by a short-range interaction. If the antiferromagnetic exchange in pnictide iron based superconductors is dominated by next neighbor coupling (J$_2$) \cite{Hu2012}, the superconducting gap can be described by the single functional form in the entire Brillouine Zone given by $\Delta(k)=\Delta_0|\cos k_x \cos k_y|$ (or $|\cos k_x + \cos k_y |$ in the 2Fe per unit cell) \cite{Hu2012b}.  Apparently, in this gap function, FS with smaller diameter would have a larger superconducting gap around Brillouin zone center, which is consistent with the measured gap size on the smallest hole pocket in several pnictide superconductors \cite{Ding2008,Lin2008,Richard2011}. At the same time, another interpretation of this feature also exists in the purely band description of the $s^{\pm}$-wave superconductivity in which the large size of the superconducting gap on the smallest pocket is attributed to the dominant interband Cooper-pair scattering between electron and hole bands of different sizes. In the absence of direct nesting the magnitude of the gap is larger on that band, which has smaller $k_{F}$\cite{Chen2015,Chubukov2016}.  In Fig. 4(f) and 4(g), we compare the measured gap sizes on different FS with $\Delta_0|\cos k_x \cos k_y|$ function. The gap size on $\alpha$ FS is much smaller than the value of the gap, expected from the fit function. ($\sim$15 meV gap size is needed in order to match.) Therefore the failure of applying  $\Delta_0| \cos k_x \cos k_y|$ function on our measured gap indicates superconductivity in CaFeK$_4$As$_4$ may not be immediately described within short-range antiferromagnetic fluctuation model. Instead, the measured gap sizes are consistent with superconductivity arising within the band description. In particular, in contrast to other ferrpnictides some of the bands in CaFeK$_4$As$_4$ are nested, {\it i.e.}- the $\beta$ FS with the largest gap size among three hole FS has the best nesting condition with $\delta$ electron band FS (Fig. 4(h)). This favors the multiband character of superconductivity in CaFeK$_4$As$_4$.}

{In conclusion, we measured the electronic structure and values of the superconducting gap of a new member of iron arsenic high temperature superconductor - CaKFe$_4$As$_4$. This material has nearly optimal T$_c$ without doping or substitution \cite{2016Meier}, but also has several hole FS sheets with significantly different diameter. This uniquely allows to study the momentum dependence of the gap magnitude in the entire Brillouine Zone and potential effects of the FS nesting present in this system. We found the superconducting gap to be isotropic within each of the FS sheets. We found strong correlation between the magnitude of the gap and nesting of the FS sheets. The larges gap is observed on $\beta$ hole and $\delta$ electron sheets, which have very similar diameters, whereas the $\alpha$ and $\gamma$ sheets have smaller values of the SC gap as they have no electron counterparts of similar diameter. This strongly supports the multiband character of the $s^{\pm}$-wave symmetry of the superconducting gap in which the Cooper-pairing forms on the electron and hole bands with strong interband repulsive interaction, enhanced by the Fermi surface nesting.}


We would like to thank Rafael Fernandes and Peter Orth for very useful discussions. This work was supported by the U.S. Department of Energy, Office of Science, Basic Energy Sciences, Materials Science and Engineering Division (sample growth, characterization and ARPES measurements). Ames Laboratory is operated for the U.S. Department of Energy by Iowa State University under contract No. DE-AC02-07CH11358. The work of F.L. and I.E. was supported by the joint DFG-ANR Grant No. ER463/8.

\bibliography{CaKFe4As4}

\end{document}